\documentclass[prd,twocolumn,showpacs]{revtex4}
\usepackage{graphicx}
\usepackage{amsmath}
\usepackage{amssymb}
\usepackage{bm}
\begin{document}

%\preprint{}
\title{Glueballs and statistical mechanics of the gluon plasma}
\author{Fabian \surname{Brau}}
\email[E-mail: ]{fabian.brau@umh.ac.be} 
\author{Fabien \surname{Buisseret}}
\thanks{F.R.S.-FNRS Postdoctoral Researcher}\email[E-mail: ]{fabien.buisseret@umh.ac.be} 
\affiliation{Groupe de Physique Nucl\'{e}aire Th\'{e}orique, Universit\'{e} de Mons, Acad\'{e}mie universitaire Wallonie-Bruxelles, Place du Parc 20, BE-7000 Mons, Belgium}
\date{\today}

\begin{abstract}
We study a pure gluon plasma in the context of quasiparticle models, where the plasma is considered as an ideal gas of massive bosons. In order to reproduce SU(3) gauge field lattice data within such a framework, we review briefly the necessity to use a temperature-dependent gluon mass which accounts for color interactions between the gluons near $T_{\text{c}}$ and agrees with perturbative QCD at large temperatures. Consequently, we discuss the thermodynamics of systems with temperature-dependent Hamiltonians and clarify the situation about the possible solutions proposed in the literature to treat consistently those systems. We then focus our attention to two possible formulations which are thermodynamically consistent and we extract the gluon mass from the equation of state obtained in SU(3) lattice QCD. We find that the thermal gluon mass is similar in both statistical formalisms. Finally, an interpretation of the gluon plasma as an ideal gas made of glueballs and gluons is also presented. The glueball mass is consistently computed within a relativistic formalism using a potential obtained from lattice QCD. We find that the gluon plasma might be a glueball-rich medium for $T\lesssim1.13\, T_{\text{c}}$ and suggest that glueballs could be detected in future experiments dedicated to the quark-gluon plasma. 
\end{abstract}
% Quark gluon plasma
%Glueball and nonstandard multiquark/gluon states
%Boson systems

\maketitle

\section{Introduction}
\label{sec:intro}
\subsection{Generalities and lattice data}
It is expected that, at high enough temperatures or densities, a phase transition from hadronic matter to quark-gluon plasma will occur. As early as 1975, Collins and Perry suggested that the dense nuclear matter at the center of neutron stars could consist in deconfined quarks and gluons~\cite{coll75}. In 1980, Shuryak studied the nuclear matter at high temperatures and introduced the terminology ``quark-gluon plasma'' in analogy with similar phenomena in atomic physics~\cite{shur80}. Beside its intrinsic interest, knowing the equation of state of a quark-gluon plasma is needed to predict the evolution of stars and to know for example if a neutron star can go through a quark phase or just collapse into a black hole. This knowledge is also important to predict when our Universe hadronized, since it is also believed that it was a quark-gluon plasma within a few $\mu$s after the Big-Bang. The hadronic matter/quark-gluon plasma phase transition, predicted by Quantum Chromodynamics (QCD), is studied experimentally at the Relativistic Heavy Ion Collider (RHIC) \cite{rhic05} and will be also studied in the future at the Large Hadron Collider (LHC). The results obtained at RHIC so far suggest that the quark-gluon plasma behaves like an almost perfect fluid instead of a weakly interacting gas, indicating that interactions are still quite large after the phase transition, for $T \gtrsim T_{\text{c}}$ ($T_{\text{c}}$ denotes the critical temperature of QCD). From a field-theoretical point of view, studying the quark-gluon plasma is a challenging task since it requires a deep understanding of QCD, and more generally of gauge theories at finite temperatures. Several frameworks have been developed and have led to a great amount of works: Perturbative methods, potential models, AdS/QCD duality, lattice QCD,\dots References about these topics can be found for example in the reviews~\cite{revs}. 

\begin{figure}[ht]
\includegraphics*[width=\columnwidth]{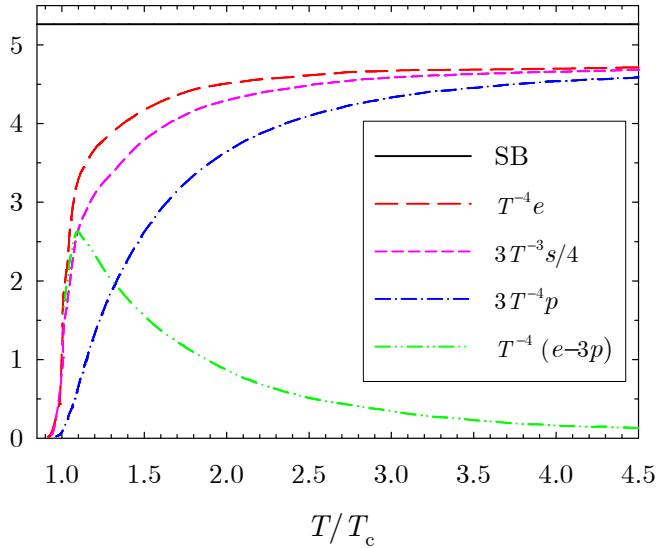}
\caption{(Color online) Energy density, entropy density, pressure, and interaction measure (or trace anomaly) of the gluon plasma versus $T/T_{\text{c}}$, as measured in pure glue lattice QCD at zero chemical potential~\cite{boyd95} (dashed lines). The full horizontal line shows the Stefan-Boltzmann limit for a gas of massless transverse gluons.}
\label{fig01}
\end{figure}

In principle, the most powerful technique to study non-perturbatively the properties of a quark-gluon plasma is lattice QCD. The equation of state of an SU(2) and SU(3) gluon plasma in lattice QCD were obtained in Refs.~\cite{enge82,enge89,boyd95,boyd,okam99}. The equation of state of a quark-gluon plasma with non-vanishing flavor number, $N_f\neq 0$, has also been computed in the more recent Refs.~\cite{kars00,alik01,aoki06,bern07}. Notice that we focus here on the case where the chemical potential vanishes, but some results have already been obtained at nonzero chemical potential (see for example Ref.~\cite{bern07b}). In Fig.~\ref{fig01}, we show the equation of state obtained from pure glue SU(3) lattice computations in the continuum limit \cite{boyd95} but in all cases ($N_f=0, 1, 2, 3$), two important features are observed: ($i$) Energy and entropy increase sharply just after the phase transition temperature while the increase for the pressure is less pronounced. ($ii$) Energy and entropy seem to saturate below the Stefan-Boltzmann constant in the range $T/T_{\text{c}}\approx2- 5$. It has been argued in Ref.~\cite{glio07} that finite-size effects were partly responsible of that behavior, but such numerical artifacts cannot explain the whole deficit. Actually, it is observed in the lattice computations of Ref.~\cite{endro} that the equation of state of the gluon plasma becomes compatible with the Stefan-Boltzman limit at very large temperatures. The ratio of pressure $p/p^{SB}$ grows for example from 0.85 around $T_{\text{c}}$ to a value compatible with 1 at $T/T_{\text{c}}= 3\, 10^7$~\cite{endro}, that is an increase with a mean slope of order 10$^{-9}$. Strictly speaking, the thermodynamical quantities do thus not \textit{saturate} below the Stefan Boltzman constant. But, since we are interested in reproducing the lattice data of Fig.~\ref{fig01} for $T/T_{\text{c}}\lesssim 5$, and because the saturation rate is so small, we will fit the lattice data in the following as if that saturation was truly realized for  $T/T_{\text{c}}\lesssim 5$. The error introduced by such an approximation is indeed completely negligible for our purpose.  

\subsection{Quasiparticle models}
There have been many attempts to understand the results obtained in lattice QCD and to derive the equation of state of a quark-gluon plasma from effective approaches. Indeed, QCD itself can be perturbatively solved only in the region of asymptotic freedom, \textit{i.e.} for very high momenta or temperatures~\cite{poli74}. But the convergence of the expansion in the strong coupling constant for the pressure is rather slow~\cite{kaja03} and consequently phenomenological models have been developed. There are mainly two frameworks: Strongly interacting quark-gluon plasma models taking explicitly into account the possible existence of bound states beyond $T_{\text{c}}$ \cite{bann95,shur04,liao05,liao05b,gelm06,gelm06b,liao06}, or quasiparticle models, where the quark-gluon plasma is described as an ideal gas of massive bosons and fermions~\cite{enge89,risc90,risc92,risc92b,golo93,pesh94,gore95,pesh96,leva98,schn01,zwan05,cast07,castor07,bann07,chandra}. In this paper, we are mainly concerned with the quasiparticle formulation of a pure gluon plasma. As we recall in Sec.~\ref{sec:sbgas}, it is necessary in such a framework to consider a phenomenological temperature-dependent gluon mass (or thermal gluon mass) in order to reproduce the lattice data. Various authors have proposed a procedure to treat statistically systems whose Hamiltonian depends on temperature: In Ref.~\cite{golo93}, the authors start with the usual partition function and simply replace the constant gluon mass, $m$, by a temperature-dependent one, $m(T)$, which leads to an invariant expression for the pressure whereas the energy and entropy are modified in order to satisfy standard thermodynamical relations between those quantities. In Ref.~\cite{gore95} however, the expression of the entropy is kept unchanged whereas both the energy and pressure are supplemented with an additional term involving $\partial_T m(T)$. Finally, in Ref.~\cite{bann07}, the author proposes to keep the expression for the energy unchanged whereas the entropy and pressure get an additional term which also involves $\partial_T m(T)$. All these three formulations are obviously not equivalent and there is still a debate to know which one, if any, is correct. In Sec.~\ref{sec:Tdepend}, we propose a way to clarify the situation starting from the first principles of statistical mechanics while in Sec.~\ref{sec:gplas} we show that all these formulations found in the literature demand similar temperature-dependent gluon masses to reproduce the lattice data. It implies that, at a qualitative level, those formulations are rather equally good. Moreover, we propose a new formulation where the expressions for the energy, entropy and pressure are invariant but where the Lagrange multiplier $\beta$ is no longer equal to $T^{-1}$ in order to ensure the so-called thermodynamic consistency, {\it i.e.} the fulfillment of the laws of thermodynamics. Note that we work in units where $\hbar=c=k_B=1$, $k_B$ being Boltzmann's constant.

Most of the quasiparticle models reach the same conclusion about the qualitative behavior of the thermal gluon mass: Just beyond the critical temperature $T_{\text{c}}$, the gluon mass has to be large and to decrease up to $T/T_{\text{c}} \simeq 1.5-2$. Then, for even larger $T$, it increases essentially linearly. We interpret the large value of the gluon mass around $T_{\text{c}}$ as a signal of strong color interactions among gluons. In Sec.~\ref{sec:glue}, we take those interactions into account by considering the gluon plasma as an ideal gas of glueballs and gluons, the color interactions being responsible for the formation of glueballs. We show that the lattice data can be reproduced if the ratio between the number of glueballs and gluons, $n(T)$, decreases monotonically when $T$ increases so that a small value is reached when the temperature is larger than the dissociation temperature of the glueball. The glueball mass and dissociation temperature are computed using a spinless Salpeter equation (to take relativistic effects into account), the potential is obtained from lattice QCD~\cite{pet05} and, to be consistent, the gluon mass is a linear function of $T$ with the same slope as the one obtained from the asymptotic analysis performed in the quasiparticle formalism. Finally, some conclusions and outlook are given in Sec.~\ref{sec:conc}.

\section{Ideal quantum gas of bosons}
\label{sec:sbgas}

\subsection{Constant mass}
As mentioned in the introduction, we study in the present work a pure gluon plasma in order to understand the available pure glue lattice data. In this section, we review briefly the reasons why an ideal gas of bosons with constant masses cannot reproduce lattice data, as well as why other attempts found in the literature, such as setting a momentum-cutoff for the gluons, are not suitable. 

From Fig.~\ref{fig01}, it is clear that the gluon plasma is not an ideal gas of massless bosons, that would only lead to the Stefan-Boltzmann constant. The first natural attempt to understand the equation of state obtained in lattice QCD is thus to consider that gluons have a constant nonzero mass. We give here the expressions of the energy density, $e_0$, entropy density, $s_0$, and pressure, $p_0$, in such a case since they will be useful later. For large enough volume, $V$, the sum over the possible quantum states is replaced by an integral and we have for a vanishing chemical potential (see for example Ref.~\cite[Chapter 5]{landau})
\begin{subequations}
\label{thermo1}
\begin{eqnarray}
	\label{energy0a}
	e_0&\equiv& \frac{E_0}{V}=\frac{d}{2\pi^2}\int_0^{\infty} dk\, k^2 q(\epsilon(k)/T)\, \epsilon(k), \\
	\label{entropy0a}
	s_0&\equiv& \frac{S_0}{V}=\frac{d}{6\pi^2 T}\int_0^{\infty} dk\, k^2 q(\epsilon(k)/T)\nonumber\\
	&&\phantom{\frac{S_0}{V}=\frac{d}{6\pi^2 T}\int_0^{\infty} dk\,}\times [k \partial_k \epsilon(k)+3 \epsilon(k)], \\
	\label{pressure0a}
	p_0&\equiv& Ts_0-e_0\nonumber\\
	&=&\frac{d}{6\pi^2}\int_0^{\infty} dk\, k^3 q(\epsilon(k)/T)\, \partial_k \epsilon(k),
\end{eqnarray}
\end{subequations}
where $d$ is the degeneracy factor, equal to 16 for transverse gluons (8 colors $\times$ 2 polarizations), and where $q(x)$ is the Bose-Einstein distribution
\begin{equation}
	\label{bose-enstein}
	q(x)=[e^{x}-1]^{-1}.
\end{equation}
Using the following dispersion relation
\begin{equation}
	\label{dispers}
	\epsilon(k)=\sqrt{k^2+m^2},
\end{equation}
Eqs.~(\ref{thermo1}) can be rewritten as
\begin{eqnarray}
	\label{energy0b}
	e_0(m,T)T^{-4}&=&\frac{d}{2\pi^2}\int_{A}^{\infty} dx\, q(x)\, x^2\, \sqrt{x^2-A^2}, \\
	              &=&\frac{d}{2\pi^2} h(A) \nonumber \\
	\label{entropy0b}
	s_0(m,T)T^{-3}&=&\frac{d}{6\pi^2}\int_{A}^{\infty} dx\, q(x)\, \sqrt{x^2-A^2}\nonumber \\
	&&\phantom{\frac{d}{6\pi^2}\int_{A}^{\infty} dx\, q(x)\,} \times \left(4x^2-A^2\right), \\
	\label{pressure0b}
	p_0(m,T)T^{-4}&=&\frac{d}{6\pi^2}\int_{A}^{\infty} dx\, q(x)\, \left(x^2-A^2\right)^{3/2},
\end{eqnarray}
where 
\begin{equation}
	A\equiv A(T)=m/T.
\end{equation}
 The so-called interaction measure (or trace anomaly) is then found to be
\begin{eqnarray}
	\label{inter}
	I(m,T)\, T^{-4}&\equiv& (e_0-3p_0)T^{-4}\nonumber\\
	&=& \frac{d}{2\pi^2} A^2 \int_{A}^{\infty} dx\, q(x)\, \sqrt{x^2-A^2}.
\end{eqnarray}
Notice that $I$ is positive (it is negative for non-relativistic bosons) and does not vanish as soon as the bosons have a nonzero mass. 

\begin{figure}[ht]
\includegraphics*[width=\columnwidth]{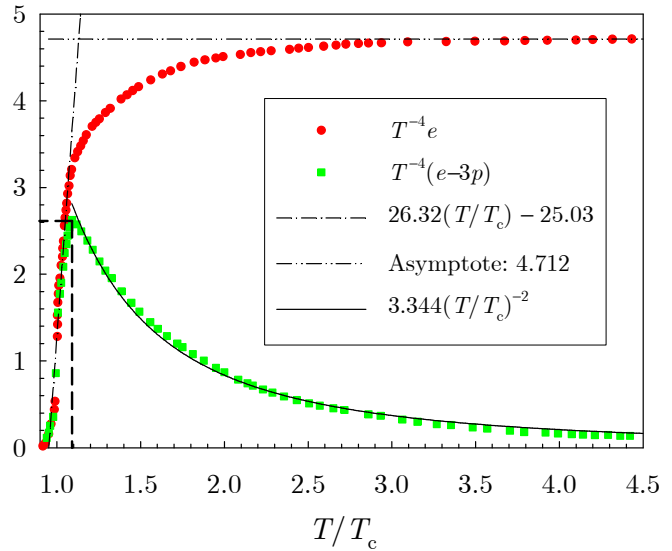}
\caption{(Color online) Energy density and interaction measure of the gluon plasma versus $T/T_{\text{c}}$, as measured in pure glue lattice QCD at zero chemical potential~\cite{boyd95} (full circles and squares). The dashed-dotted lines outline the apparent saturation value at $T/T_{\text{c}}\lesssim 5$ and the behavior near $T_{\text{c}}$ of the energy density. The solid line is a fit of the interaction measure beyond its maximal value (dashed line).   }
\label{fig02}
\end{figure}

This one-parameter -- $m$ -- model cannot reproduce the lattice data even if there is some qualitative agreement. In particular, it cannot reproduce simultaneously some of the important characteristics summarized in Fig.~\ref{fig02}:
\begin{enumerate}
	\item For $T/T_{\text{c}}\lesssim 5$, the energy density apparently saturates around 4.712, which is about $10\%$ below the Stefan-Boltzmann constant $d\, \pi^2/30=5.264$ (the gap is even of about $20 \%$ when one considers $N_f \neq 0$ \cite{kars00}),
	\item Around the critical temperature, the energy density increases very fast with a mean slope roughly equal to 26,
	\item The maximum of the interaction measure is located at $T/T_{\text{c}}\simeq 1.1$ and its value is about 2.6,
	\item The decay of the interaction measure is given in good approximation by $3.344\, (T/T_{\text{c}})^{-2}$.
\end{enumerate}
Indeed, a model with constant mass predicts that energy density, entropy density, and pressure will quickly saturate at the Stefan-Boltzmann constant. It is already enough to discard the model if we consider that the saturation occurs at the smaller value given above, namely 4.712 instead of 5.264. However, even if we consider that lattice data did not yet saturate, the following arguments show that a model with a constant gluon mass is not able to reproduce these data since their four features enumerated above imply four very different values of the ratio $m/T_{\text{c}}$. At $T/T_{\text{c}}=4.5$, the value of the energy (\ref{energy0b}) is equal to the lattice value 4.712 if $A\simeq 1$, which implies $m/T_{\text{c}}\simeq 4.5$. But the slope of the energy (\ref{energy0b}) can be bounded from above:
\begin{eqnarray}
	\partial_{T/T_{\text{c}}}(e_0 T^{-4})&=&\frac{d}{2\pi^2}(-\partial_{A} h(A)) (-\partial_{T/T_{\text{c}}} A) \\
	&\le & \frac{d}{2\pi^2} 1.6 \frac{m}{T_{\text{c}}} \left(\frac{T}{T_{\text{c}}}\right)^{-2}.
\end{eqnarray}
Consequently, around $T_{\text{c}}$, the derivative of the energy density is smaller than around $1.3\, m/T_{\text{c}}$, which implies $m/T_{\text{c}}\simeq 20$ to reproduce the data. The interaction measure (\ref{inter}) has a maximum located at $A=2.303$ and the value of the maximum is equal to $1.2 d/(2\pi^2)\simeq 1$ (far from 2.6 obtained in lattice QCD). To reproduce the position of the maximum obtained from lattice data we find that $m/T_{\text{c}}\simeq 2.5$. At last, the decay of the interaction measure predicted by this model for large enough temperature is $d/12 A^2$. Again to reproduce the data, we need $m/T_{\text{c}}\simeq 1.6$. It is a simple exercise to verify %along the same lines 
that even with a degeneracy factor, $d$, considered as a free constant parameter, one cannot get a good quantitative agreement between this model and the lattice data in the whole available temperature range.

\subsection{Momentum cutoff}

Another version of this simple model is obtained by introducing a cutoff, $K$, for small momenta in Eqs.~(\ref{thermo1}) as in Refs.~\cite{risc92,risc92b}. The physical motivation of such a cutoff is that, near the critical temperature, gluons with low momenta should be bound into glueballs and should thus not contribute to thermodynamical quantities related to an ideal gas of gluons. It is easy to see that this modified model predicts also a quick saturation of the energy density, entropy density, and pressure at the Stefan-Boltzmann constant. Moreover the additional parameter $K$ does not help to describe the large value of the derivative of the energy density around $T_{\text{c}}$. This can be seen by using arguments similar to the ones given above when $K=0$ or by looking at the various figures of Refs.~\cite{risc92,risc92b}. Another possibility is to consider a cutoff, $K(T)$, which depends on the temperature as in Ref.~\cite{enge89}. To have a relevant physical meaning, $K(T)$ should be a decreasing function of $T$ since one expects that, for high enough temperatures, glueballs will not be present in the plasma (see also Sec.~\ref{sec:glue}). Using arguments similar to the ones developed to get the general form of $m(T)$ in Sec.~\ref{ssec:genem}, one can prove that indeed, around $T_{\text{c}}$, we have $\partial_T K(T)<0$. However, to reproduce the saturation of the thermodynamical quantities below the Stefan-Boltzmann constant, the cutoff must increase linearly with $T$ for large enough temperatures. Qualitatively, the shape of $K(T)$ would be similar to the one of $m(T)$, see Fig.~\ref{fig2}. This behavior of $K(T)$ for large $T$ is problematic and its physical meaning is not obvious.

\subsection{Temperature-dependent mass}

For the reasons mentioned above, various authors have considered that a temperature-dependent gluon mass is the most relevant ingredient to be added to this model~\cite{golo93,pesh94,gore95,pesh96,leva98,schn01,castor07,bann07}. Indeed, since energy density, entropy density, and pressure are decreasing functions of $A$, a simple way to make them saturate below the Stefan-Boltzmann constant for large $T$ is to have a mass $m(T)$ such that $A=m(T)/T$ saturates to a non-vanishing constant in this regime of temperature. This implies that one must have 
\begin{equation}\label{constm}
	m(T) \sim T \quad \text{for} \quad T\gg T_{\text{c}}.
\end{equation}

Although $m(T\gtrsim T_{\text{c}})$ is mostly a phenomenological parameter that has to be fitted in order to reproduce the lattice results, the thermal gluon mass can be related to another important parameter characterizing a plasma, that is the plasma frequency. In a QED plasma for example, photons cannot propagate with a frequency below the plasma frequency. The situation is similar in the quark-gluon plasma: Gluons (plasmons) cannot propagate as free particles if their energy is too low. In fact, gluons acquire a thermal mass which is the plasma frequency. Perturbative calculations confirm that point: To leading order, the thermal gluon mass, also proportional to the Debye mass at large temperatures~\cite{leva98}, has been found to be proportional to $\sqrt{\alpha_s(T)}\ T$~\cite{shur78,kapu79}. Following standard notation, $\alpha_s(T)=g^2(T)/4\pi$ with $g(T)$ the strong coupling constant. Higher-order contributions have been calculated in the literature; it appears that the perturbative expansion converges very slowly and thus that the leading-order result is only valid at very high temperatures~\cite{kaja97}. 
Since $\sqrt{\alpha_s(T)}\sim1/\sqrt{\ln(T)}$ varies more slowly than $T$, the linear behavior of the thermal gluon mass is found to be dominant at very large $T$ in perturbative QCD. The constraint~(\ref{constm}) is thus in good qualitative agreement with already known results. 

It is worth mentioning that a resummation of the leading-order formulas leads to a modification of the thermal gluon mass that improves the convergence of the results toward the Stefan-Boltzman limit. See in particular Ref.~\cite{blai}, focusing on the convergence of the pressure, and Ref.~\cite{rebh} in which the hard thermal loop perturbation theory is successfully applied to reproduce lattice data at finite chemical potential within a quasiparticle approach. Higher-order perturbative calculations also give good results~\cite{laine}. 

%Introducing a temperature-dependent mass is not a trivial problem since standard statistical mechanics only deals with Hamiltonians which do not explicitly depend on temperature. As we pointed out in the introduction, there is a debate in the literature about how statistical mechanics should be modified in such a case. In the next section we propose a way to clarify the current situation and give an alternative solution with respect to what can be found in the literature.

\section{Statistical mechanics with temperature-dependent Hamiltonians}
\label{sec:Tdepend}

In a phenomenological model describing the gluon plasma, it is quite natural to assume that such a plasma can be seen as some gas of quasiparticles, or quasigluons, as many authors have done in previous works (see the introduction). Standard statistical mechanics is the necessary framework to deal with gases of quasigluon. But, as we stressed in the previous section, a qualitative description of the pure glue lattice data demands the introduction of a temperature-dependent mass for the quasigluons. Such a thermal mass also emerges from QCD itself, as shown within perturbation theory. 

The problem is that standard statistical mechanics only deals with Hamiltonians which do not explicitly depend on temperature. There is a debate in the literature about how it should be modified in such a case to be consistent. In this section, we propose an extension of statistical mechanics aiming at treating temperature-dependent Hamiltonians. Our procedure does not only allow to recover other existing formalisms in a unified way, it also leads to another new formulation which preserves the standard definition of energy, entropy, and pressure. We first consider classical systems in equilibrium in the canonical ensemble and restrict our study to reversible processes to keep the discussion as simple as possible. Then the case of an ideal quantum gas of bosons, relevant for our study, is presented. 

\subsection{General formalism}
\label{ssec:genef}

Let us consider a probability density which is a function of the Hamiltonian $H(p_i,q_i,T)$: $\rho\equiv\rho(H)$. $\rho$ is the density of probability of finding the dynamical variables of the system, namely $\{p_i,q_i\}$, within some volume of the phase space. We assume here that the Hamiltonian depends explicitly on the temperature $T$. The form of the function $\rho$ can be determined using standard procedures; we give here the main ideas and refer the reader to standard textbooks for more details like, for example, Ref.~\cite[p.50-55]{wale00}. Consider two subsystems $A$ and $B$ with Hamiltonians $H_A$ and $H_B$ at equilibrium. The probabilities to find the subsystem $A$ within the phase-space volume $d\lambda_A$ and $B$ within $d\lambda_B$ are given respectively by $dP_A=\rho_A(H_A)\, d\lambda_A$ and by $dP_B=\rho_B(H_B)\, d\lambda_B$ following the definition of the probability density. Consider then the system $C$ obtained by combining the two subsystems $A$ and $B$. The probability to find both $A$ within $d\lambda_A$ and $B$ within $d\lambda_B$, hence to find $C$ within $d\lambda_C=d\lambda_A\, d\lambda_B$, is given by $dP_C = dP_A\, dP_B = \rho_A(H_A)\, \rho_B(H_B)\, d\lambda_C$. If the system $C$ is also at equilibrium, we can write that $dP_C =\rho_C(H_C)\, d\lambda_C$. Neglecting the interaction between $A$ and $B$, we have that $H_C=H_A+H_B$ and
\begin{equation}
	\rho_C(H_A+H_B)=\rho_A(H_A)\, \rho_B(H_B).
\end{equation}
The intuitive idea underlying this constraint is that the description of a given system as a whole or as composed of two subsystems at equilibrium must be equivalent. Such a functional equation has a unique non trivial solution:
\begin{eqnarray}
	\rho_A(x)&=&C_1 e^{-\beta x}, \quad \rho_B(x)=C_2 e^{-\beta x}, \nonumber \\\text{and}\quad \rho_C(x)&=&C_1 C_2 e^{-\beta x},
\end{eqnarray}
where $C_1, C_2$, and $\beta$ are arbitrary constants with respect to the dynamical variables. Consequently, the general form of the normalized probability density is
\begin{eqnarray}
	\label{norm-prob}
	\rho(H)&=&\frac{e^{-\beta H}}{\int e^{-\beta H} d\lambda}, \\
	&\equiv&\frac{e^{-\beta H}}{{\cal Z}(\beta)},
\end{eqnarray}
where we have introduced the partition function ${\cal Z}$ although it does no longer play a central role in the present formulation. For the moment, $\beta$ is still arbitrary and will be fixed later. We recall that $\int\, d\lambda$ denotes an integration on the phase space of the system.

Another elegant way to obtain the expression~(\ref{norm-prob}) for the probability density is to use the so-called maximum-entropy estimate. According to Jaynes~\cite{jayn57}, this is the least biased possible estimate on the available information. He showed that the form~(\ref{norm-prob}) maximizes the entropy of the probability distribution, \textit{i.e.} the Shannon's measure \cite{shan48}, given by
\begin{equation}
	\label{entropy-def}
	S\equiv -\overline{\ln \rho}=-\int \rho \ln \rho\, d\lambda,
\end{equation}
where we have introduced the notation $\overline{x}$ for the phase-space average of the quantity $x$. Notice that, in this formulation, $\beta$ is a Lagrangian multiplier and, as above, does not depend on the dynamical variables. One advantage of this derivation is that it shows straightforwardly the intimate link between entropy and probability density. The next quantity we consider is the energy, defined as the averaged Hamiltonian
\begin{equation}
	\label{energy-def}
	E\equiv \overline{H}=\int H \rho\, d\lambda.
\end{equation}

The only unknown in the above relations is $\beta$. We assume that this parameter depends only on the temperature $T$ in the following way:
\begin{equation}
	\label{fbeta-def}
	T=\frac{1}{f(\beta)}.
\end{equation}
The function $f(x)$ in assumed to be positive, monotonic and such that $f(x)\in[0,\infty[$ so that the function $\beta(T)$ can be unambiguously defined. In order to determine $f(\beta)$, we use the laws of thermodynamics, that give a relation between energy, entropy and $f(\beta)$.  The first law links the variation of internal energy, $E$, of heat, $Q$, and of work, $W$, as follows: $dE =\delta Q +\delta W$. For reversible processes we have $\delta W=-p\, dV$. The second law relates the variation of heat with the variation of entropy: $dS=\delta Q /T$ (the equality holds for reversible processes only). Assuming that both the internal energy and the entropy are functions of $\beta$ -- thus implicitly of $T$ -- and of $V$, and combining the first and the second laws of thermodynamics, we obtain
\begin{equation}
	\partial_{\beta}E\, d\beta+\partial_V E\, dV = T(\partial_{\beta} S\, d\beta +\partial_V S\, dV) - p\, dV.
\end{equation}
Equating the terms in $d\beta$ and $dV$, we find a relation between energy and entropy
\begin{equation}
	\label{link-energy-entropy}
	\partial_{\beta} S = f(\beta)\, \partial_{\beta} E,
\end{equation}
where we used Eq.~(\ref{fbeta-def}) and, in the thermodynamical limit, the following expression for the pressure
\begin{equation}
	\label{pression-def}
	p \equiv T s -e = (TS-E)/V.
\end{equation}
Equation (\ref{link-energy-entropy}) is the relation that allows to determine $f(\beta)$ and to know the relation between $\beta$ and $T$ through Eq.~(\ref{fbeta-def}). Indeed, substituting expression (\ref{norm-prob}) into the definition (\ref{entropy-def}) of the entropy and using Eq.~(\ref{energy-def}), we obtain
\begin{equation}
	S=\ln {\cal Z} + \beta E.
\end{equation}
The derivation with respect to $\beta$ of the entropy leads to
\begin{eqnarray}
	\label{deriv-entropy}
	\partial_{\beta} S &=& \frac{\partial_{\beta} {\cal Z}}{{\cal Z}} + E + \beta \partial_{\beta} E, \nonumber \\
	&=& -\beta\, \overline{\partial_{\beta} H} + \beta \partial_{\beta} E.
\end{eqnarray}
A comparison between Eqs.~(\ref{link-energy-entropy}) and (\ref{deriv-entropy}) gives an equation for $f(\beta)$:
\begin{equation}
	\label{fbeta}
	f(\beta)=\beta\left[1- \frac{\overline{\partial_{\beta} H(p_i,q_i,T=1/f(\beta))}}{\partial_{\beta} E(T=1/f(\beta))} \right].
\end{equation}
This is, in general, a nonlinear first order differential equation for $f$. One thus gets $f(\beta,c)$ and, thanks to Eq.~(\ref{fbeta-def}), the function $\beta(T,c)$. The integration constant $c$ can be constrained (sometime even fixed) by imposing that $f(\beta)$ is positive, bounded and monotonic for $\beta \in [0,\infty[$. Moreover, if there exists a temperature $T_0$ such that $\left.\partial_T H\right|_{T=T_0}=0$, the boundary condition needed to determine uniquely the solution of this equation is obtained by imposing that $\beta(T_0,c)=1/T_0$ in order to recover the usual formalism at this particular temperature. To clarify the procedure, we give an explicit example in Appendix~\ref{app:Tdepend}.

In this approach, the entropy, energy, and pressure are given respectively by their usual expressions~(\ref{entropy-def}), (\ref{energy-def}), and (\ref{pression-def}). But, the dependence on $T$ of the Hamiltonian enforces a particular link between $\beta$ and $T$, that can be found through the resolution of the nontrivial relation~(\ref{fbeta}). In standard problems, $\partial_T H=0$ and one recovers the well-known link $f(\beta)=\beta=1/T$ as a solution of Eq.~(\ref{fbeta}). This general procedure has the serious advantage of preserving the formal expressions of all the relevant thermodynamical quantities of the problem: The only modification arises at the level of the definition of $\beta$, which is not a physical parameter in itself. For computational applications however, this formalism is rather complicated since it needs an \textit{a priori} knowledge of the solution of Eq.~(\ref{fbeta}) if one wants to extract physical informations about the system under study. Equation (\ref{fbeta}) can only be solved once the dependence on temperature of the Hamiltonian is explicitly known. However in the context of gluon plasma, the thermal mass of the gluons is unknown and must be determined from lattice data. That is why it is of interest to find more tractable ways of dealing with temperature-dependent Hamiltonians. As we show in the following, our procedure allows to find such formulations, that corresponds to frameworks already in use in the literature. The study of the formalism developed in this section will be the subject of a forthcoming paper. 

\subsection{Alternative solutions}
\label{ssec:alter}

First, notice that Eq.~(\ref{deriv-entropy}) can be rewritten as
\begin{equation}
	\label{new-entropy}
	\partial_{\beta} \tilde{S}\equiv \partial_{\beta}( S+ \int_{\beta_{\star}}^{\beta} \nu\, \overline{\partial_{\beta} H}|_{\beta=\nu} \, d\nu) = \beta \partial_{\beta} E,
\end{equation}
where we have introduced a new form, $\tilde{S}$, for the entropy and where $\beta_{\star}$ is some integration constant. The relation between the new entropy and the energy is then formally identical to the one given by Eq.~(\ref{link-energy-entropy}) provided we choose $f(\beta)=\beta$ as in standard statistical mechanics. The modified pressure is given by Eq.~(\ref{pression-def}) where $S$ is replaced by $\tilde{S}$. In this formulation, the standard expressions for the energy and for $\beta$ are preserved but the expressions for the entropy and pressure are modified. We then loose the usual connection between the probability density and the entropy. This formalism has been proposed in Ref.~\cite{bann07}.

Second, it is readily observed that another equivalent rewriting of Eq.~(\ref{deriv-entropy}) is
\begin{equation}
	\label{new-energy}
	\partial_{\beta}S = \beta \partial_{\beta} (E- \int_{\beta_{\star}}^{\beta} \overline{\partial_{\beta} H}|_{\beta=\nu} d\nu) \equiv \beta \partial_{\beta} \tilde{E},
\end{equation}
where we have introduced a new form, $\tilde{E}$, for the energy. Again, the relation between the new entropy and the energy is formally identical to the one given by Eq.~(\ref{link-energy-entropy}) provided that we choose also $f(\beta)=\beta$. The modified pressure is given by Eq.~(\ref{pression-def}) where $E$ is replaced by $\tilde{E}$. In this formulation, the standard expressions for the entropy and for $\beta$ are preserved but the expressions for the energy and pressure are modified. In particular, the energy is no longer the average of the Hamiltonian. This formalism was first proposed in Ref.~\cite{gore95} and used in several other works, for example in Refs.~\cite{pesh96,leva98,schn01,rebh}. 

It is worth mentioning a third procedure that has been used in Ref.~\cite{golo93} and consists in preserving the expression of the pressure. We mention it for completeness but will not further study it in the present work. In can be deduced from Eq.~(\ref{pression-def}) that 
\begin{equation}
p=\frac{T}{V}\, \ln{\cal Z}. 	
\end{equation}
Consequently, one can leave the pressure invariant by setting $f(\beta)=\beta$ and by computing ${\cal Z}$ as usually done. But in this case, the entropy and the energy will be modified since the laws of thermodynamics demand that $e=s/\beta -p$, with $s=-\beta^2\partial_\beta p$: a term in $\partial_\beta H$ appears because of the $\partial_\beta p$ term.

These three alternative solutions are derived from the laws of thermodynamics where the usual link $f(\beta)=\beta$ is kept, but each one requires the standard form of the thermodynamical quantities to be modified. We think that the formalism developed in the previous section is the most fundamental one since it only demands a redefinition of $\beta$, which is only a Lagrangian multiplier. Moreover, this redefinition of $\beta$ as a function of $T$ is only local since this is done through the differential equation (\ref{fbeta}): $1/T\equiv f(\beta)=\beta$ when $\partial_T H(p_i,q_i,T)=0$. In contrast, the corrections to the thermodynamic quantities obtained from the alternative formulations derived in this section are non-local; they involve an integral over some range of temperatures, see Eqs.~(\ref{new-entropy}) and (\ref{new-energy}). This implies that even if $\partial_T H(p_i,q_i,T)=0$ in some large interval of temperatures, the corrections can be quite significant if the integration is performed over an interval of $T$ for which the Hamiltonian depends on $T$. However, the formalism proposed in Sec.~\ref{ssec:genef} is far more complicated to deal with in numerical, phenomenological, applications when the dependence on temperature of the Hamiltonian is not known a priori. That is why the other approaches are useful too: From a computational point of view, it is easier to compute the extra integrals appearing in Eqs.~(\ref{new-entropy}) or (\ref{new-energy}) than to solve Eq.~(\ref{fbeta}) if the temperature dependence of the Hamiltonian on $T$ is not known. 

\subsection{Ideal quantum gas of bosons}

The derivations presented in Secs.\ref{ssec:genef} and \ref{ssec:alter} also holds formally in the quantum case provided that the average takes correctly into account the statistics of bosons and fermions. We now focus on the case we are eventually interested in: a gluon plasma. As we did in the classical case, we start with the standard expressions for the energy density, entropy density, and pressure, \textit{i.e.} 
\begin{subequations}\label{thermo0}
\begin{eqnarray}
	\label{energy0c}
	e_0&=&\frac{d}{2\pi^2}\int_0^{\infty} dk\, k^2 q(\beta \epsilon(k,\beta))\, \epsilon(k,\beta), \\
	\label{entropy0c}
	s_0&=&\frac{\beta d}{6\pi^2}\int_0^{\infty} dk\, k^2 q(\beta \epsilon(k,\beta))\nonumber\\
	&&\phantom{\frac{\beta d}{6\pi^2}\int_0^{\infty} dk\, k^2 }\times [k \partial_k \epsilon(k,\beta)+3 \epsilon(k,\beta)], \\
		\label{pressure0c}
	p_0&=&\frac{d}{6\pi^2}\int_0^{\infty} dk\, k^3 q(\beta \epsilon(k,\beta))\, \partial_k \epsilon(k,\beta),
\end{eqnarray}
\end{subequations}
where $\epsilon(k,\beta)\equiv \epsilon(k,1/f(\beta))$. We actually assume that the temperature dependence of $\epsilon$ is known, the relation between $T$ and $\beta$ being given by Eq.~(\ref{fbeta-def}) which was obtained from the laws of thermodynamics. 

It is convenient to write the entropy as follows
\begin{equation}
	\label{entropy0d}
	s_0=\ln {\cal Z}_0 + \beta e_0,
\end{equation}
where
\begin{equation}
	\ln {\cal Z}_0=-\frac{d}{2\pi^2}\int_0^{\infty}dk\, k^2\ln(1-e^{\beta \epsilon(k,\beta)}).
\end{equation}
The derivative of $s_0$ with respect to $\beta$ leads to
\begin{equation}
	\label{deriv-entropy2}
	\partial_{\beta} s_0 = -\beta \overline{\partial_{\beta} \epsilon}+ \beta \partial_{\beta} e_0,
\end{equation}
where the average is now given by
\begin{equation}
	\label{new-average}
	\overline{\partial_{\beta} \epsilon} = \frac{d}{2\pi^2}\int_0^{\infty}dk \, k^2 q(\beta \epsilon(k,\beta))\partial_{\beta} \epsilon(k,\beta).
\end{equation}
The comparison between Eq.~(\ref{deriv-entropy2}) and Eq.~(\ref{link-energy-entropy}) leads to the following expression for $f(\beta)$:
\begin{equation}
	\label{fbeta2}
	f(\beta)= \beta\left[1- \frac{\overline{\partial_{\beta} \epsilon(T=1/f(\beta))}}{\partial_{\beta} e_0(T=1/f(\beta))} \right].
\end{equation}
This formula for $f$ is formally identical to the one obtained in the classical case; only the definition of the average is different. The same comments about the boundary condition can be made. 

It is also possible to recover in the quantum case the other formalisms proposed in the literature. In the formulation where the expression of the energy density is preserved, the expression for the new entropy density is still given by Eq.~(\ref{new-entropy}) but the average is now defined by Eq.~(\ref{new-average}) and the pressure can be obtained through Eq.~(\ref{pression-def}). A similar remark applies for the formalism which preserves the expression of the entropy, see Eq.~(\ref{new-energy}). Consequently, what we call Model 1 in the next sections is defined by the equations
\begin{subequations}\label{model1}
\begin{equation}
	\label{mod1}
	e^{(1)} = e_0, \quad 	s^{(1)} = s_0+B^{(1)}, \quad	p = p_0+\frac{B^{(1)}}{\beta},
\end{equation}
with
\begin{equation}
	\label{b1}
	B^{(1)}(\beta)=\int_{\beta^{(1)}_{\star}}^{\beta} \nu\, \overline{\partial_{\beta} \epsilon}|_{\beta=\nu} d\nu.
\end{equation}
\end{subequations}
Model 2 is similarly defined by
\begin{subequations}\label{model2}
\begin{equation}
	\label{mod2}
	e^{(2)} = e_0-B^{(2)}, \quad	s = s_0,\quad	p = p_0+B^{(2)},
\end{equation}
with
\begin{equation}
	\label{b2}
	B^{(2)}(\beta)=\int_{\beta^{(2)}_{\star}}^{\beta} \overline{\partial_{\beta} \epsilon}|_{\beta=\nu} d\nu.
\end{equation}
\end{subequations}
In both Models 1 and 2 we have $\beta=1/T$ as usual, and $e_0$, $s_0$, and $p_0$ are given by Eqs.~(\ref{thermo0}). The integration constants, $\beta^{(1)}_{\star}$ and $\beta^{(2)}_{\star}$, are arbitrary parameters. The function $\epsilon(k,\beta)$ is given by
\begin{equation}
\label{disper}
\epsilon(k,\beta)=\epsilon(k,1/T)=\sqrt{k^2+m^2(T)},
\end{equation}
with $m(T)$ the thermal gluon mass. If that function was unambiguously known, Models 1 and 2 would lead to inequivalent results and the general solution~(\ref{fbeta2}) should rather be used. However, $m(T)$ is a parameter of the model, that can be fitted on some lattice data. Its behavior at large $T$ should nevertheless be coherent with Eq.~(\ref{constm}), in agreement with perturbative QCD. Thus it can be expected that all the presented formalisms should lead to very similar results provided that the gluon mass and the integration constants, $\beta^{(i)}_\star$, are properly chosen. In particular, Models 1 and 2 are compared in Sec.~\ref{sec:gplas}, where we show that they lead to similar temperature-dependent gluon masses and that they are both able to reproduce the lattice data with excellent accuracy. Consequently, from a practical point of view, they are both rather equivalent because there is not much constraints on this thermal gluon mass.

\section{Application to the gluon plasma}
\label{sec:gplas}

\subsection{General features of $m(T)$}
\label{ssec:genem}

In this section, we give some general characteristics of the function $m(T)$. Let us consider the Model 1 where the expression of the energy is conserved (similar conclusion can be obtained with Model 2 by using the expression of the entropy). From the lattice data, we know that the slope of the energy near the critical temperature is positive and rather large (see Fig.~\ref{fig02}). From Eq.~(\ref{energy0b}) we find
\begin{equation}
	\partial_{T/T_{\text{c}}}(e_0 T^{-4})=\frac{d}{2\pi^2}(\partial_A h(A)) (\partial_{T/T_{\text{c}}} A) \ge 0.
\end{equation}
Since $\partial_A h(A)\le 0$, we find that $m(T)/T$ is a decreasing function of the temperature
\begin{equation}
	\partial_{T/T_{\text{c}}} \frac{m(T)}{T} \le 0.
\end{equation}
From Eq.~(\ref{energy0b}), it is easy to see that in order to reproduce the saturation of the energy for $T/T_{\text{c}} \gtrsim 4$, the thermal mass $m(T)$ should be a linear function of $T$: $m(T)=\bar{m}\, T$, {\it i.e.} $A=\bar{m}$, with $\bar{m}=0.973$. We can also show that the large mean slope of the energy near the critical temperature implies that the derivative of $m(T)$ should be negative in this region. Indeed,
\begin{equation}
	\partial_{T/T_{\text{c}}}(e_0 T^{-4})\le \frac{d}{2\pi^2}\max(-\partial_A h(A)) \max(-\partial_{T/T_{\text{c}}} A).
\end{equation}
From the expression (\ref{energy0b}) of $h(A)$, it is readily computed that $\max(-\partial_A h(A))=M\simeq 1.6$. Let us assume that $\partial_T m(T)\ge 0$ everywhere, then
\begin{eqnarray}
	\max(-\partial_{T/T_{\text{c}}} A)&=&\max\left(\frac{m(T)}{T_{\text{c}}} \left(\frac{T_{\text{c}}}{T}\right)^2-\frac{T_{\text{c}}}{T}\partial_Tm(T)\right), \nonumber \\
	&\le& \max\left(\frac{m(T)}{T_{\text{c}}} \left(\frac{T_{\text{c}}}{T}\right)^2 \right) \nonumber \\
	&\le& \max\left(\frac{\tilde{m}(T)}{T_{\text{c}}} \left(\frac{T_{\text{c}}}{T}\right)^2 \right),
\end{eqnarray}
where $\tilde{m}(T)=\bar{m} T$ for $T>\tilde{T}$ and $\tilde{m}(T)=\bar{m} \tilde{T}$ for $T<\tilde{T}$ and where $\tilde{T}$ is the temperature where the thermal mass should be linear ($\tilde{T}/T_{\text{c}} \sim 2-3$). Consequently we obtain
\begin{eqnarray}
	\partial_{T/T_{\text{c}}}(e_0 T^{-4})|_{T/T_{\text{c}}\simeq 1}&\le& \left.\frac{d}{2\pi^2}M \bar{m} \frac{\tilde{T}}{T_{\text{c}}} \left(\frac{T_{\text{c}}}{T}\right)^2\right|_{T/T_{\text{c}}\simeq 1} \nonumber \\
	&\lesssim& 1.3 \frac{\tilde{T}}{T_{\text{c}}} \lesssim 4,
\end{eqnarray}
since $T/T_{\text{c}}\simeq 1$, $\tilde{T}/T_{\text{c}} \simeq 2-3 $, $\bar{m}\simeq 1$ and $M \simeq 1.6$. This upper bound on the derivative of the energy is much lower than the value given by the lattice data. Large values for this derivative can only be obtained if $\partial_T m(T)< 0$ near the critical temperature.

In consequence, we have analytically shown that an agreement with lattice QCD can be obtained provided that 
\begin{equation}\label{manaly}
	\partial_T m(T\gtrsim T_{\text{c}})<0,\quad{\rm and}\quad  m(T\gg T_{\text{c}})=0.973\, T.
\end{equation}

\subsection{Numerical results}
\label{ssec:fitm}

In the previous section, we have shown that the shape of $m(T)$ is rather constrained by the lattice data within the frameworks of Models 1 and 2. Let us now explicitly extract numerically $m(T)$ from these data. 

\begin{figure}[b]
\includegraphics*[width=\columnwidth]{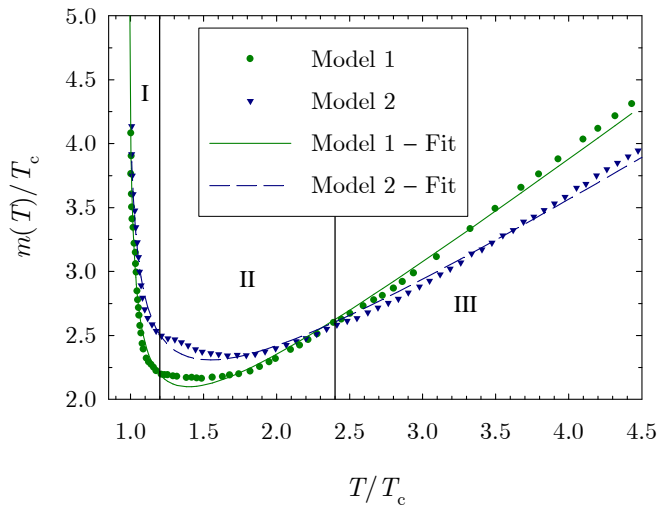}
\caption{(Color online) Thermal gluon masses obtained by fitting Model 1 (circles) and Model 2 (triangles) to the lattice data of Ref.~\cite{boyd95}, see Fig.~\ref{fig01}. Models 1 and 2 are defined by Eqs.~(\ref{model1}) and (\ref{model2}) respectively, with the dispersion relation~(\ref{disper}). The fitted forms~(\ref{mfit1}) and (\ref{mfit2}) are also plotted (solid lines).}
\label{fig2}
\end{figure}

We begin by considering Model 1, where the form of the energy density is preserved and consequently given by Eq.~(\ref{energy0c}) in which the dispersion relation~(\ref{disper}) is chosen. At a given temperature, $T^*$, the thermal gluon mass, $m(T^*)$, can be obtained by numerically solving the equation $e^{(1)}(T^*,m(T^*))=e_0(T^*,m(T^*))=e^{{\rm lat}}(T^*)$, where $e^{{\rm lat}}(T^*)$ is the lattice energy density at the considered temperature. The computed gluon mass is plotted in Fig.~\ref{fig2} and can be well fitted by the following form
\begin{subequations}\label{mfit1}
\begin{equation}\label{mmodel1}
	\frac{m^{(1)}(T)}{T_{\text{c}}}=m_0\, \frac{T}{T_{\text{c}}}+\frac{m_1}{(T/T_{\text{c}}-m_2)^{m_3}},
\end{equation}
with
\begin{eqnarray}
	m_0&=&0.873,\quad m_1=0.612,\nonumber\\
	m_2&=&0.983,\quad m_3=0.411.
\end{eqnarray}
\end{subequations}

The observation of Fig.~\ref{fig2} and of the fitted form~(\ref{mfit1}) clearly shows the different behaviors predicted in Sec.~\ref{ssec:genem}. First, the linear increase of $m(T)$ is obvious for $T/T_{\text{c}}\geq 2.5$, and corresponds to the region III in Fig.~\ref{fig2}. However, the slope $m_0$ differs from the asymptotic value of $0.973$ predicted in the previous section by about 10\%. This can be understood by remarking that $m_3$ is rather small while $m_1$ is of the same order of magnitude than $m_0$: The term supplementing the linear one in Eq.~(\ref{mmodel1}) still brings a non negligible contribution at large temperature, causing the fitted slope $m_0$ to be smaller than in the case of a genuine linearly rising mass, see Eq.~(\ref{manaly}). Second, $m(T)$ strongly decreases for $T/T_{\text{c}}\simeq 1.0-1.2$, corresponding to region I in Fig.~\ref{fig2}. The fitted form we get is actually singular near the critical temperature, the parameter $m_3$ playing the role of a critical exponent. Third, there exists an intermediate zone between the singular and the linear behaviors, in which $m(T)$ reaches a minimum. This zone corresponds to region II in Fig.~\ref{fig2}.    

In Model 2, the form of the energy density is no longer preserved as in Model 1, but the entropy density is so. That is why the thermal gluon mass at a given temperature, $T^*$, can be numerically computed in a very similar way by solving the equation $s^{(2)}(T^*,m(T^*))=s_0(T^*,m(T^*))=s^{{\rm lat}}(T^*)$. We recall that $s_0$ is given by Eq.~(\ref{entropy0c}). The computed thermal gluon mass is plotted in Fig.~\ref{fig2} and can be accurately fitted by the following form  
\begin{subequations}\label{mfit2}
\begin{equation}\label{mmodel2}
	\frac{m^{(2)}(T)}{T_{\text{c}}}=k_0\, \frac{T}{T_{\text{c}}}+\frac{k_1}{(T/T_{\text{c}}-k_2)^{k_3}},
\end{equation}
with
\begin{eqnarray}
	k_0&=&0.724,\quad k_1=0.982,\nonumber\\
	k_2&=&0.973,\quad k_3=0.345.
\end{eqnarray}
\end{subequations}
Equation~(\ref{mfit2}) is formally equivalent to (\ref{mfit1}); only the values of the numerical coefficients are slightly different. Thus the same comments as for Model 1 can be done. It is worth noting that the regions where the linear increase and strong decrease occur are identical within Models 1 and 2. A physical interpretation of the thermal gluon mass can thus be given independently of the considered Model. 
\begin{figure}[ht]
\includegraphics*[width=\columnwidth]{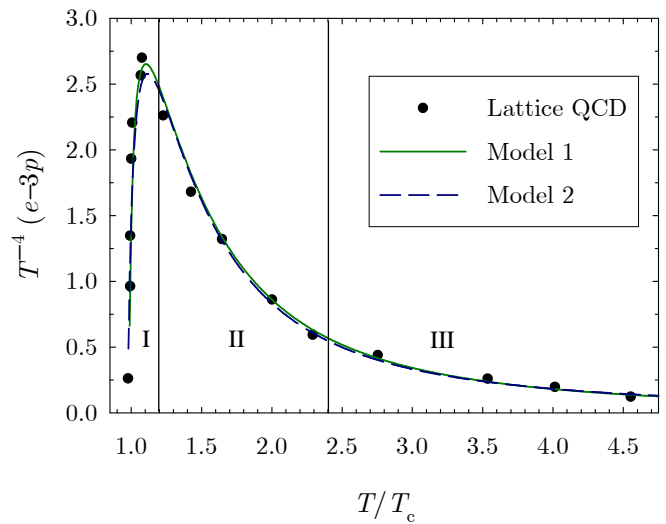}
\caption{(Color online) Interaction measure of the gluon plasma computed with Model 1 (solid line) and Model 2 (dashed line), and compared to the lattice data of Ref.~\cite{boyd95} (circles). Model 1 is defined by Eqs.~(\ref{model1}) with the gluon mass~(\ref{mfit1}) and Model 2 is defined by Eqs.~(\ref{model2}) with the gluon mass~(\ref{mfit2}). The dispersion relation~(\ref{disper}) and the values~(\ref{betas}) for the integration constants are used.}
\label{fig3}
\end{figure}
\begin{figure}[ht]
\includegraphics*[width=\columnwidth]{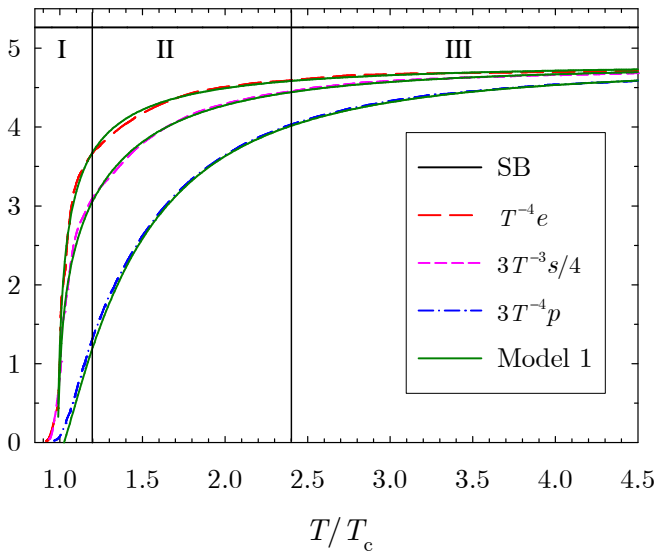}
\caption{(Color online) Same as Fig.~\ref{fig01}, but the results obtained with Model 1 are also plotted for comparison. Model 1 is defined by Eqs.~(\ref{model1}) with the gluon mass~(\ref{mfit1}), the dispersion relation~(\ref{disper}), and the value~(\ref{betas}) for the integration constant.}
\label{fig4}
\end{figure}

The thermal gluon mass has been fitted on one of the three thermodynamical quantities available in lattice QCD: energy density for Model 1, and entropy density for Model 2. The remaining quantities can now be numerically computed within both Models by using Eqs.~(\ref{model1}) and (\ref{model2}) with the dispersion relation~(\ref{disper}), provided that the integration constants ensuring an optimal agreement with lattice QCD are known. The following fitted values
\begin{equation}\label{betas}
T_{\text{c}}\, \beta^{(1)}_\star=0.435,\quad {\rm and}\quad T_{\text{c}}\, \beta^{(2)}_\star=0.445,
\end{equation}
lead to an excellent agreement with the available lattice data, as it can be seen in Figs.~\ref{fig3} and \ref{fig4}. 

The interaction measures computed with Model 1 and Model 2 are quasi indistinguishable from each other and from the lattice data. Again, both formalisms lead to nearly identical results. That is why, for clarity, we have only plotted the results of Model 1 in Fig.~\ref{fig4}: The curves computed with Model 2 would have been indistinguishable from those of Model 1. Notice that in our approach, the thermal gluon mass is fitted so that the asymptotic behavior of the interaction measure corresponds to lattice QCD, that is $e-3p\propto T^2$ (see Fig.~\ref{fig02}). Such a quadratic increase is compatible with previous theoretical results~\cite{pisa06} and with the more recent unquenched lattice study~\cite{chen07}. It is worth mentioning that other approaches rather favor $e-3p\propto T$~\cite{zwan04,mill06,giac07}. The interaction measure is thus a quantity that would deserve further studies since there is not yet a general agreement concerning its asymptotic growing.

The values we find for the integration constants are almost equal: Their average value is $T_{\text{c}}\, \bar\beta=0.44$, corresponding to the temperature $\bar T/T_{\text{c}}\simeq 2.27$. This is the typical temperature at which $m^{(1)}(\bar T)=m^{(2)}(\bar T)$, as it can be seen in Fig.~\ref{fig2}. It is not a coincidence: Models 1 and 2 are designed to reproduce the same data. Then if $m^{(1)}(\bar T)=m^{(2)}(\bar T)$, the only way for both Models to give identical results is to have $B^{(i)}(\bar\beta)=0$, thus $\beta^{(1)}_\star=\beta^{(2)}_\star=\bar \beta$, in rough agreement with the fitted values~(\ref{betas}). The integration constant is thus not really a free parameter since its value can constrained by the thermal gluon mass once Model 1 and Model 2 are compared.

Finally, it is important to stress that the terms $B^{(i)}$, given by Eqs.~(\ref{b1}) and (\ref{b2}), are not small corrections as one could have thought. Without these terms, even the qualitative behavior of the various thermodynamical quantities is wrong. 

\subsection{Color interactions above $T_{\text{c}}$}
\label{ssec:scenar}

We have considered up to now that the gluon plasma is an ideal boson gas, where gluons are transverse and free but have a temperature-dependent mass $m(T)$. From Sec.~\ref{ssec:genem} we can conclude that reproducing the lattice QCD results demands first that $m(T\gg T_{\text{c}})\sim T$ and second that $m(T\gtrsim T_{\text{c}})$ decreases fast enough (see Fig.~\ref{fig2}). Such a nontrivial behavior can be intuitively explained by invoking color interactions above $T_{\text{c}}$. It is indeed widely accepted that, at the critical temperature $T_{\text{c}}$, the medium undergoes a phase transition and becomes deconfined. It does not mean however that the color interactions vanish: They are actually screened because of the great amount of color charges in the medium, and the residual potential is no longer confining as for $T=0$. These residual color interactions can be rather important for $T/ T_{\text{c}}\simeq1-2$, as suggested by several lattice QCD studies~\cite{pet05}. 

One can think about the mean field approximation to have a first guess about the influence of screened color interactions. In this picture, the gluon dispersion relation should be modified as follows:
\begin{equation}
\label{disper2}
	\sqrt{k^2+m^2(T)}=\sqrt{k^2+\bar m^2(T)}+\bar V(T),
\end{equation}
where $\bar V(T)$ is the effective mean potential energy felt by a gluon, and where $\bar m(T)$ is \textit{a priori} different of $m(T)$. When $T$ becomes very large, it can reasonably be assumed that $\bar V(T)$ vanishes. Consequently, one has $\bar m(T\gg T_{\text{c}})=m(T\gg T_{\text{c}})=\bar m\, T$. Our main physical assumption is then the following: Since $\bar m(T)$ is the thermal gluon mass in a temperature range when the gluons are free, it can be seen as the rest mass of a free gluon for any $T\geq T_{\text{c}}$. Let us now consider that $\bar V(T\gtrsim T_{\text{c}})\gg \bar m(T\gtrsim T_{\text{c}})$, \textit{i.e.} that the color interactions become dominant near $T_{\text{c}}$. Then, by squaring Eq.~(\ref{disper2}), one gets $m^2(T)=\bar m^2(T)+\bar V^2(T)+2\sqrt{k^2+\bar m^2(T)}\bar V(T)$ and consequently $m^2(T\gtrsim T_{\text{c}})\approx \bar V^2(T\gtrsim  T_{\text{c}})$ since the potential term dominates the right hand side.

We are thus led to the following interpretation for $m(T)$: At large $T$ (region III in Fig.~\ref{fig2}), $m(T)$ tends to the rest mass of a free gluon in the gluon plasma according to the perturbative QCD result. But near $T_{\text{c}}$ (region I in Fig.~\ref{fig2}), the behavior of $m(T)$ is dominated by the existence of non negligible screened color interactions. Region II in Fig.~\ref{fig2} is finally a transition regime in which these interactions progressively vanish. According to this scenario, screened color interactions play an important role for $T/T_{\text{c}}\simeq1-2$, and one can wonder whether glueballs can form or not at these temperatures. The next section is devoted to answer to this question. Notice that the possible glueball formation above $T_{\text{c}}$ has already been suggested in Ref.~\cite{castor07} as a mechanism explaining the sudden increase of the effective gluonic degrees of freedom near $T_{\text{c}}$ that is observed in this last work.

\section{Existence of glueballs above $T_{\text{c}}$}
\label{sec:glue}

\subsection{Effective Hamiltonian for glueballs}
\label{ssec:hamglu}

In a constituent gluon (or quasiparticle) picture such as the one we develop here, a glueball is a bound state of at least two gluons. Let us focus on two-gluon glueballs. Being the lightest and presumably the most strongly bound ones, they should be the easiest glueballs to be produced in the gluon plasma. In a deconfined medium, a binary gluon state may exist in several colored configurations following the decomposition $\bm 8\otimes\bm 8=\bm 1\oplus\bm 8\oplus\bm 8\oplus\bm{10}\oplus\overline{\bm{10} }\oplus\bm{27}$. As the strength of color interactions is proportional to the color Casimir operator of the gluon pair, the last three configurations are irrelevant as far as glueball formation is concerned since they lead to interactions which are either vanishing ($\bm{10},\ \overline{\bm{10}}$) or repulsive $\bm{27}$~\cite{shur04,boul08}. However, both the singlet and octet configurations lead to attractive interactions, these interactions in the singlet channel being twice as large as in the octet one. The most favored glueball from an energetic point of view is thus a two-gluon bound state with the gluon pair in a color singlet. The dynamics of the gluon pair also comes into play at this stage: The most strongly bound gluon pairs will be those with a minimal value of the radial quantum number ($n=0$) and of the orbital angular momentum. If the gluons were longitudinal the minimal value of the square orbital angular momentum would be $\left\langle \bm L^2\right\rangle=0$ for the $0^{++}$ state. However, we have seen that the large-$T$ behavior of the gluon plasma is compatible with transverse gluons. In this case, $\left\langle \bm L^2\right\rangle=2$ is the minimal allowed value, corresponding to the $0^{\pm+}$ glueballs as shown in Ref.~\cite{math08}. 

Denoting the static potential between a color-singlet quark-antiquark pair by $V(r,T)$, a relativistic Hamiltonian describing the aforementioned lightest glueballs is the following spinless Salpeter one
\begin{equation}\label{hamglu}
	H_G=2\left.\sqrt{\bm p^2+\bar m^2(T)}\right|_{\left\langle \bm L^2\right\rangle=2}+\frac{9}{4}\, V(r,T),
\end{equation}
where $\bm p^2=p^2_r+\left\langle \bm L^2\right\rangle/r^2$ and where
\begin{equation}
\frac{\bar m(T)}{T_{\text c}}=\bar m\, \frac{T}{T_{\text c}}	=0.973\, \frac{T}{T_{\text c}}
\end{equation}
is the free gluon mass introduced in the previous section. The $9/4$ factor comes from the color Casimir operator. Such a Casimir scaling for the static energy between sources in various color representation has been confirmed by the lattice study~\cite{colo}. We choose for $\bar m$ the value that reproduces that saturation value of the thermodynamical quantities following the analysis of Sec.~\ref{ssec:genem}. Notice that, since $T_{\text c}$ is estimated to be around $270$~MeV by pure glue lattice calculations~\cite{boyd}, one has
\begin{equation}
\bar m(T)=0.263\, \frac{T}{T_{\text c}}~\ {\rm GeV}.
\end{equation}

We point out that building an effective glueball Hamiltonian by starting from a best-known quark-antiquark one has already led to a successful description of glueballs at $T=0$~\cite{math08}. That is why we find relevant to apply it in this case also. A last remark has to be done: Our framework leads by construction to the same mass for the scalar and pseudoscalar glueballs. At $T=0$ this degeneracy can be lifted by the introduction of instanton-induced forces~\cite{math08}. Such forces are not taken into account by the present model and one can expect that the ground state of Hamiltonian~(\ref{hamglu}), whose mass is denoted $M_G(T)$, is rather an average mass of the scalar and pseudoscalar glueballs. Although the current understanding of this topic is far from being complete, we can nevertheless mention that instantons effects might be less important at high temperatures following Ref.~\cite{shur96}. Actually, we have checked that the results that we obtain in the following do not demand an accurate knowledge of $M_G(T)$.

\begin{figure}[ht]
\includegraphics*[width=\columnwidth]{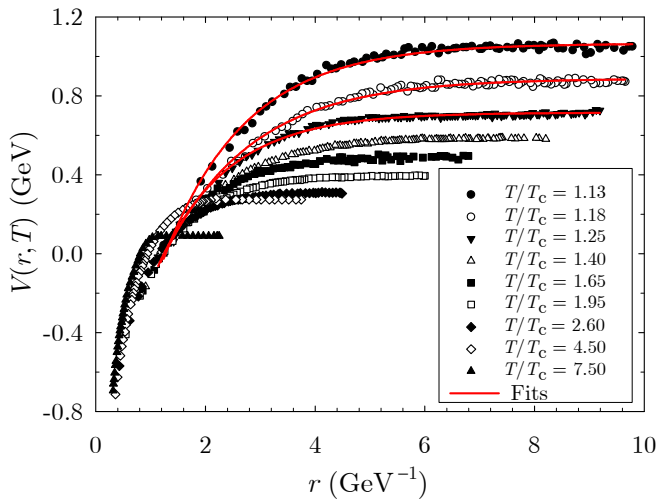}
\caption{(Color online) Internal energy of a static quark-antiquark pair computed in lattice QCD for different values of $T/T_{\text{c}}$ and for $N_f=0$ (dots). Lattice data are taken from Refs.~\cite{pet05} and compared to the fitted form~(\ref{Vfit}) for some values of $T/T_{\text{c}}$ (solid lines).}
\label{fig5}
\end{figure}

A key ingredient in Hamiltonian~(\ref{hamglu}) is the potential energy $V(r,T)$ between a static quark-antiquark pair. It is well known from lattice QCD that this potential is compatible with a funnel shape $ar-b/r$ at $T=0$~\cite{bali00}, but the situation is less clear when $T>0$. The potential energy that is the most readily obtained in lattice QCD is the quark-antiquark free energy $F(r,T)$~\cite{mcler,boyd}. We recall that, thermodynamically speaking, the free energy of a system is the energy that is available in the system to produce a work once the energy losses due to the increase of the entropy have been subtracted. As also noticed in Ref.~\cite{shur04}, in a potential approach however, the potential energy of the system should be the total energy that it contains, no matter it will be lost or not in heat transfers. Such a potential energy corresponds to the internal energy of the system, usually denoted by $U=F+TS$, where $S$ is the entropy. The internal energy of a quark-antiquark pair is thus the quantity we choose as potential term. It has been computed in lattice QCD in Refs.~\cite{pet05}; we give a plot of these results in Fig.~\ref{fig5}. Notice that those $N_f=0$ computations are the most relevant for our purpose since we consider a genuine gluon plasma. 

\begin{figure}[ht]
\includegraphics*[width=\columnwidth]{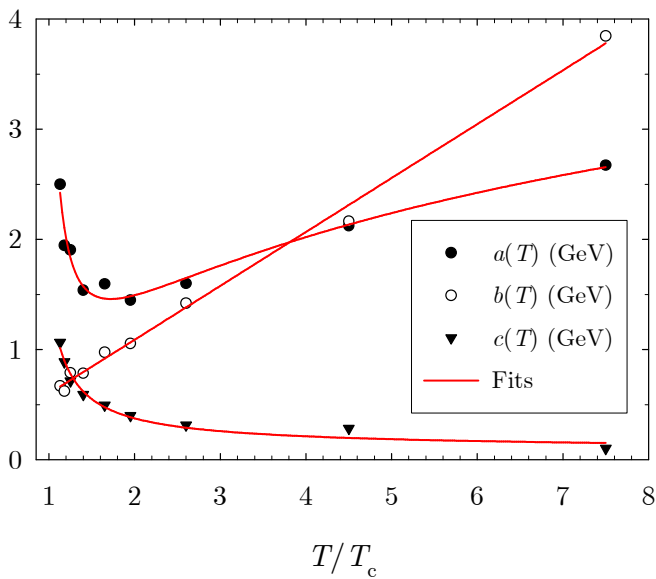}
\caption{(Color online) Values of $a(T)$, $b(T)$, and $c(T)$ obtained by a fit of the lattice QCD data to the form (\ref{vfit2}) (dots), compared to the analytical curves (\ref{fitcoeff}) (solid lines).}
\label{fig6}
\end{figure}

It can be checked in Figs.~\ref{fig5} and \ref{fig6} that the lattice data are accurately fitted by the following form
\begin{subequations}\label{Vfit}
\begin{equation}\label{vfit2}
	V(r,T)=-a(T)\ {\rm e}^{-b(T)\, r}+c(T),
\end{equation}
where 
\begin{eqnarray}
\label{fitcoeff}
	a(T)&=&\frac{a_0}{\ln\left(\frac{T}{a_1T_{\text{c}}}\right)} + a_2 \ln\left(\frac{T}{a_1T_{\text{c}}}\right), b(T)= b_0+ b_1\, \frac{T}{T_{\text{c}}},\nonumber \\ 
	c(T)&=&\frac{c_0}{\ln\left(\frac{T}{c_1T_{\text{c}}}\right)},\quad {\rm and}
\end{eqnarray}
\begin{eqnarray}
 a_0&=&0.459~{\rm GeV},\quad a_1=0.915,\quad a_2=1.159~{\rm GeV}	,\nonumber\\
  b_0&=&0.111~{\rm GeV},\quad b_1=0.489~{\rm GeV}, \nonumber \\
   c_0&=&0.341~{\rm GeV},\quad  c_1=0.808.
\end{eqnarray}
\end{subequations}
It has to be stressed that the form~(\ref{Vfit}) is the one that gives the best fit of the lattice data but is not motivated by any physical theory predicting such a form. We just use it as a convenient parameterization of the lattice QCD results. It clearly appears from Eq.~(\ref{vfit2}) that the potential energy is no longer confining. Exponential potentials indeed only admit a finite number of bound states.   

\subsection{Numerical results}
\label{ssec:numres} 

\begin{figure}[t]
\includegraphics*[width=\columnwidth]{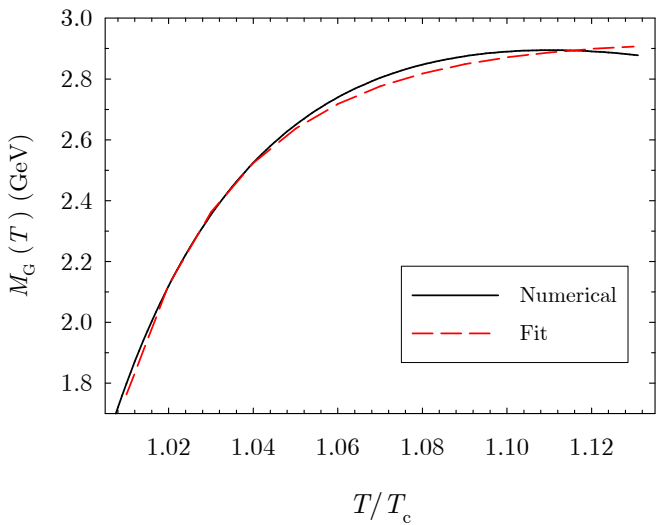}
\caption{(Color online) Numerically computed lowest-lying glueball mass, that is the ground state mass of Hamiltonian~(\ref{hamglu}), versus $T/T_{\text{c}}$ (solid line). The fitted form~(\ref{mfit}) is also plotted for comparison (dashed line). The curve stops at the glueball dissociation temperature, namely 1.13\, $T_{\text{c}}$.}
\label{fig7}
\end{figure}

All the terms appearing in Hamiltonian~(\ref{hamglu}) are now explicitly known, and its ground state mass can be numerically computed. To this aim we use the Lagrange mesh method, which is a numerical procedure allowing in particular to accurately solve eigenequations associated to relativistic Hamiltonians~\cite{sem01}. The evolution of the lowest-lying glueball mass with the temperature is given in Fig.~\ref{fig7}. The numerically computed evolution of the glueball mass with $T$ is accurately fitted by the form
\begin{subequations}\label{mfit}
\begin{equation}\label{mgit2}
	M_G(T)=\frac{9}{4}c(T)+2 \bar m(T)+2b(T)\, \varepsilon(T)\quad {\rm for}\quad T\leq 1.13\, T_{\text{c}} ,
\end{equation}
where 
\begin{equation}\label{mfit3}
\varepsilon(T)=\frac{\varepsilon_0 T/T_{\text{c}}-\varepsilon_1}{T/T_{\text{c}}-\varepsilon_2},
\end{equation}
\begin{equation}\label{mfit4}
	\varepsilon_0=0.818,\quad \varepsilon_1=0.921,\quad \varepsilon_2=0.958.
\end{equation}
\end{subequations}
The glueball mass we find is around $1.8$~GeV at $T\approx \, T_{\text{c}}$, then increases to reach a maximal value of about 2.8~GeV. Notice that the mass near the critical temperature is similar to the one obtained at zero temperature~\cite{glub}. To our knowledge, the behavior of the scalar glueball mass versus the temperature has not been studied a lot in the literature. We can nevertheless quote the lattice study of Ref.~\cite{glub2} that finds a reduction of 20\% of the scalar glueball mass when one goes from $T=0$ to $T=T_{\text{c}}$, and the more recent work~\cite{meng} finding an almost constant glueball mass from $T=0$ to $T=T_{\text{c}}$. Beyond the qualitative behavior of $M_G(T)$, an important result we find is that the ground state is bound up to $T=1.13\, T_{\text{c}}$ and then dissociates in the medium above this temperature. Numerically, the dissociation temperature is reached when the binding energy of the system vanishes. Our model thus predicts the existence of glueballs in the temperature range $T/T_{\text{c}}=1-1.13$, but the existence of bound states is a very stringent criterion: Glueball resonances can indeed appear in the continuum even if the gluons are not bound. Following the lattice results of Ref.~\cite{meng}, glueball resonances can even be expected up to $1.9\, T_{\text{c}}$.  

Since glueballs can be present in the deconfined medium, we propose to recompute the thermodynamical properties of the gluon plasma by assuming that it is a mixing between an ideal gas of transverse gluons and an ideal gas of glueballs; the glueball abundance $n(T)$ depending on the temperature. This last approach will be referred to as Model 3; it shares with Model 1 the property that $\beta=1/T$ and that the form of the energy is preserved. Its spirit is a bit similar to the one of the hadronic resonance gas model, assuming that the hot hadronic medium can be described as an ideal gas made of all
possible resonance species (see Refs.~\cite{hgm,hgm2} for more informations). We actually consider that the screened color interactions ``generate" color singlet scalar and pseudoscalar glueballs in a first stage, and that these glueballs behave as free particles in the gluon plasma in a second stage. Consequently, if 
\begin{equation}
e_0(d,m,\beta)=\frac{d}{2\pi^2} \int^\infty_{\beta m} \frac{k^2\sqrt{k^2-(\beta m)^2}}{{\rm e}^k-1}\, dk 
\end{equation}
is the energy density of an ideal gas of bosons with mass $m$ and with $d$ degrees of freedom, then the total energy density of the mixed gluon-glueball gas is
\begin{subequations}\label{model3}
\begin{eqnarray}
\label{emodel3}
	e^{(3)}&=&[1-n(T)]\, e_0(16,\bar m T,1/T)\nonumber\\
	&&+n(T)\, e_0(2,M_G(T),1/T),
\end{eqnarray}
where two degrees of freedom are associated to the glueball gas, accounting for the lowest-lying $0^{\pm+}$ states. \textit{A priori}, $n(T)$ should vanish above $1.13\, T_{\text{c}}$ because glueballs are then not bound anymore. However, two-gluon resonances can in principle appear in the continuum above the dissociation temperature. The simplest way to take this phenomenon into account is to allow $n(T)$ to be nonzero above the dissociation temperature. In this sector, formula~(\ref{mgit2}) remains well-defined, roughly simulating a gluon pair in the continuum. 
 \begin{figure}[t]
\includegraphics*[width=\columnwidth]{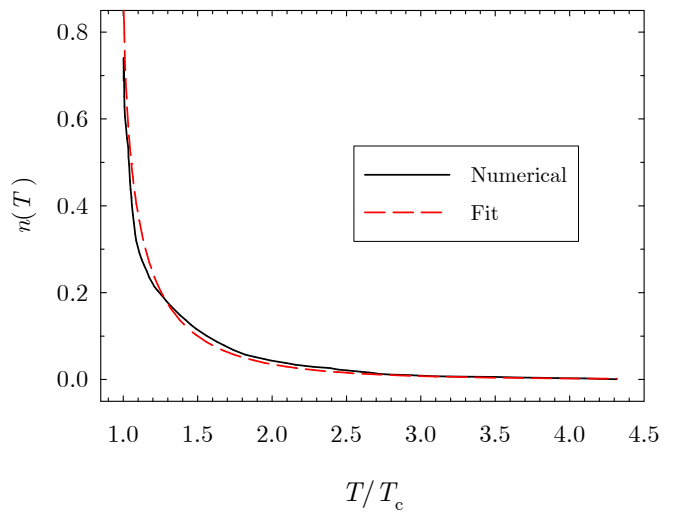}
\caption{(Color online) Glueball abundance computed by fitting Eq.~(\ref{emodel3}) to lattice QCD versus $T/T_{\text{c}}$ (solid line). The fitted form~(\ref{nt}) is plotted for comparison (dashed line).}
\label{fig8}
\end{figure}  

The unknown function $n(T)$ can be computed by fitting Eq.~(\ref{emodel3}) to the lattice energy. The result is given in Fig.~\ref{fig8}; it appears that the numerically computed curve is accurately described by the following form
\begin{equation}
\label{nt}
	n(T)={\rm e}^{-n_0\, (T/T_{\text{c}} -1)^{n_1}},
\end{equation}
with $n_0=3.358$ and $n_1=0.541$. The glueball abundance is nearly $100$\% at $T=T_{\text{c}}$,  then decreases to reach 33\% at 1.13 $T_{\text{c}}$, the dissociation temperature of the two-gluon glueballs. But as we said previously, resonances are then still expected to form in the continuum, justifying a nonzero glueball abundance at higher temperatures. Finally, $n(T)$ is less than 5\% at 1.9 $\ T_{\text{c}}$. Such a negligible value is coherent with the fact that glueball resonances are expected to disappear above that temperature~\cite{meng}. We have checked that the quantitative behavior of $n(T)$ is not very sensitive to the glueball mass, $M_G(T)$. The key result of Model 3 is rather that $e_0(16,\bar m T,1/T)$ alone is unable to fit the available data and consequently that an additional term accounting for glueballs is needed.  

The entropy density can be computed from Eqs.~(\ref{emodel3}) and (\ref{link-energy-entropy}). It reads
\begin{equation}\label{smodel3}
	s^{(3)}=\int^{1/T_{\text{c}}}_{1/T}\beta\, \partial_\beta e^{(3)}(1/\beta)\, d\beta.
\end{equation}
The upper bound of this last integral ensures that $s^{(3)}(T_{\text{c}})=0$, in qualitative agreement with lattice QCD. Finally, the pressure can be computed thanks to the definition~(\ref{pression-def}), that is 
\begin{equation}\label{pmodel3}
p^{(3)}=T\, s^{(3)}-e^{(3)}	.
\end{equation}
\end{subequations}
The results are plotted in Figs.~\ref{fig9} and \ref{fig10} and compared to lattice QCD. As it was the case for Models 1 and 2, Model 3 leads to an excellent agreement with the lattice data, although relying on a different physical picture of the gluon plasma.

\begin{figure}[ht]
\includegraphics*[width=\columnwidth]{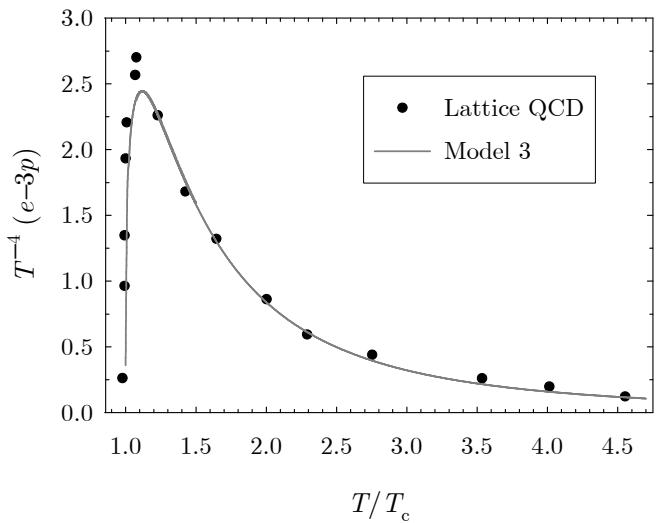}
\caption{Same as Fig.~\ref{fig3}, but lattice data are this time compared to Model 3 (solid gray line) defined by Eqs.~(\ref{model3}).}
\label{fig9}
\end{figure}

\begin{figure}[ht]
\includegraphics*[width=8.0cm]{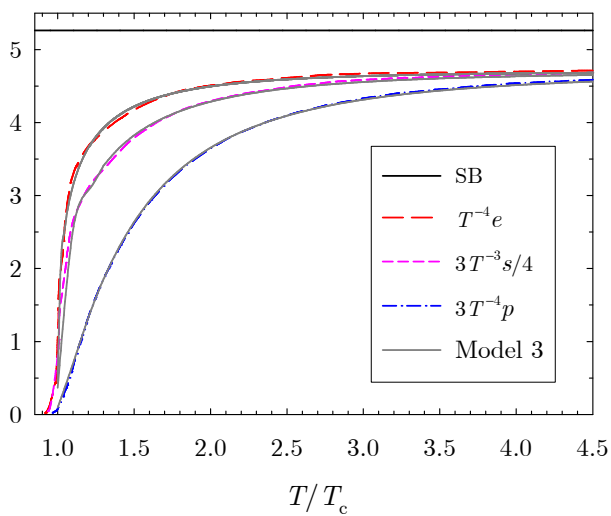}
\caption{(Color online) Same as Fig.~\ref{fig4}, but lattice data are this time compared to Model 3 (solid gray line) defined by Eqs.~(\ref{model3}).}
\label{fig10}
\end{figure}

\section{Conclusions and outlook}
\label{sec:conc}

It is now widely accepted that the equation of state of the gluon plasma, coming from pure gauge lattice QCD computations, can be accurately reproduced by modeling the gluon plasma as a gas of transverse gluons with a temperature-dependent mass. As we have outlined in the beginning of this paper indeed, such a quasiparticle model is indeed in disagreement with lattice QCD if a constant gluon mass is used. One is thus led to deal with temperature-dependent Hamiltonians. In that case, standard formulas in statistical mechanics have to be modified in order to enforce the thermodynamical consistency, but the procedure to achieve such a task varies from one work to another. In the frameworks that can be found in the literature so far, the standard expression of only one thermodynamical quantity can be preserved in order to enforce the thermodynamic consistency, \textit{i.e.} to satisfy the laws of thermodynamics. The expressions of the other quantities have to be modified: Either the pressure~\cite{golo93}, the entropy~\cite{gore95}, or the energy~\cite{bann07} is kept invariant. 

In this work, we have clarified the situation by showing that all the existing formulations can be derived in a simple unified way. In the process, we have uncovered a new possible formulation for which the standard form of each thermodynamical quantity is preserved but for which $\beta$ is no longer equal to $1/T$. The function $\beta(T)$ has to be extracted from a first order nonlinear differential equation expressing the fulfillment of the laws of thermodynamics. We think that this last formalism is the most fundamental one, since it only demands a change in the definition of the Lagrangian multiplier $\beta$, which has no physical meaning \textit{a priori}. Moreover, the corrections to standard statistical mechanics implied by this new formulation are only local in $T$ -- {\it i.e.} they vanish in regions where the Hamiltonian does not depend on $T$ -- while corrections found in other formulations are non-local in $T$. However, this new formulation is far more complicated to deal with in numerical applications when the dependence of the Hamiltonian on temperature is not known. That is why the other approaches are also useful to study the quark-gluon plasma.  

Consequently, we focused on two formulations: The ones that preserve the form of the energy and of the entropy. It can be analytically shown that, independently of the considered formulation, reproducing the lattice data leads to constraints on the thermal gluon mass, $m(T)$. It must be strongly decreasing just after the critical temperature and grow linearly asymptotically. A numerical fit of the thermal mass on the available data confirms this behavior and eventually leads to an excellent agreement with lattice QCD. Both frameworks lead to nearly indistinguishable results as expected, and to very similar thermal gluon mass. 

Mean-field-inspired arguments show that the singular behavior of the thermal gluon mass near $T_{\text{c}}$ accounts for residual color interactions, which are still strong in the early stages after deconfinement. The potential energy coming from such screened color interactions has already been computed in lattice QCD, allowing us to build a consistent Hamiltonian describing the interactions between two transverse gluons in a color singlet, that is the channel in which the color interactions are the strongest. It appears that the two-gluon ground state, corresponding to the scalar and pseudoscalar glueballs, remains bound up to $T=1.13\, T_{\text{c}}$. We have then proposed a last description of the gluon plasma, in which this medium is seen as a ideal mixture of free gluons and colorless glueballs. The agreement with lattice QCD is as good as with the previous approaches, with a glueball abundance that is very large near the critical temperature, takes the lower value of 33\% at the dissociation temperature of the lightest glueballs, and becomes negligible after $1.9\, T_{\text{c}}$, where even continuum glueball resonances are expected to disappear~\cite{meng}.  This interpretation of the gluon plasma draws a bridge between the quasiparticle approach and other models focusing on the existence of bound states after deconfinement~\cite{shur04}. 

From an experimental point of view, the main result of the present study is the prediction that the gluon plasma, and thus presumably the quark-gluon plasma, might be a glueball-rich medium in the early stages after deconfinement. This brings support to previous studies arguing that an important amount of glueballs can be formed in relativistic heavy ion collisions~\cite{gluprod,gluprod2}. The experimental detection of the scalar glueball in the quark-gluon plasma could be achieved through the scenario developed in Refs.~\cite{vento06} which roughly suggests that, although the bare scalar glueball would be nearly stable in the quark-gluon plasma, it should mix with scalar mesons. Then, such a ``physical" glueball, denoted as $G$, could decay mostly in the channels $G\rightarrow\pi\pi$ and $G\rightarrow\gamma\gamma$ through its mesonic component, leading to an enhancement of the number of events versus the two-photon (or two-pion) invariant mass. In our model, the bare glueball mass is mostly located around $2.8$~GeV; a peak in the $\gamma\gamma$ or $\pi\pi$ channels can thus be expected not too far of $2.8$~GeV, depending on the strength of the meson-glueball coupling.

We finally stress that, if quarks were included in our model, the number of bound states above the critical temperature would increase since mesons, diquarks, quark-gluon states, \textit{etc.} can also form. We leave the extension of our approach to the full quark-gluon plasma for future works. The effects of a nonzero chemical potential will also be leaved for subsequent studies.

\begin{acknowledgments}
The authors thank Vincent Mathieu, Francesco Giacosa and O. Kaczmarek for useful discussions and suggestions about the present work. F. Buisseret thanks the F.R.S.-FNRS Belgium for financial support. F. Brau acknowledges financial support from a return grant delivered by the Federal Scientific Politics.
\end{acknowledgments}

\begin{appendix}
\section{Determination of $\beta(T)$ for temperature-dependent Hamiltonians: an example}
\label{app:Tdepend}
In order to illustrate the general procedure given in Sec.~\ref{ssec:genef}, we study the particular case of a classical ideal gas with temperature-dependent mass $m(T)$, or equivalently $m(f(\beta))$ because of the definition~(\ref{fbeta-def}). The Hamiltonian reads
\begin{equation}
	H=\frac{k^2}{2m(f(\beta))},
\end{equation}
and one finds that, for system of $N$ particles, 
\begin{equation}
	{\cal Z}=V^N\left[\frac{2\pi m(f(\beta))}{\beta}\right]^{3N/2},
\end{equation}
where $V$ is the volume of the system. The normalized probability density is thus known and it can be computed that 
\begin{equation}
 E=\frac{3N}{2\beta},\quad {\rm and} \quad	\overline{\partial_\beta H}=-\frac{3N}{2\beta}\frac{m'(f(\beta)) f'(\beta)}{m(f(\beta))},
\end{equation}
where the prime denotes a partial derivation with respect to the argument of the considered function. These last two equalities allow to rewrite Eq.~(\ref{fbeta}) as 
\begin{equation}\label{deq1}
	\beta\frac{m'(f(\beta)) }{m(f(\beta))}f'(\beta)+f(\beta)-\beta=0.
\end{equation}

Let us consider the following form for $m(T)$ to illustrate the procedure
\begin{equation}
	m(T)=m_0 e^{(1-\sqrt{1+4\delta/T})/2},
\end{equation}
with $\delta\ge 0$. The equation for $f(\beta)$ then reads
\begin{equation}
	\label{tempf}
	\delta \beta^2 f'(\beta)=(f(\beta)-\beta) \sqrt{1+4 \delta f(\beta)}.
\end{equation}
To determine uniquely the function $f$, we need a boundary condition. For $T\to \infty$, the mass tends to $m_0$ which is constant. In this limit, one recover the standard statistical mechanics and $f(\beta)=\beta=1/T$. Consequently the boundary condition is $f(0)=0$. The unique solution of the nonlinear differential equation (\ref{tempf}) is then
\begin{equation}
	f(\beta)=\beta + \delta \beta^2.
\end{equation}
The relation between $\beta$  and $T$ is then given by
\begin{equation}
	\beta=\frac{-1+\sqrt{1+4\delta/T}}{2\delta}.
\end{equation}
For $T\gg 4\delta$ (or $\delta \to 0$), we just recover the standard relation $\beta=1/T$. In this formalism, we also find that
\begin{equation}
	\frac{E}{N}=\frac{3\delta}{-1+\sqrt{1+4\delta/T}}.
\end{equation}

\begin{figure}[t]
\includegraphics*[width=8.0cm]{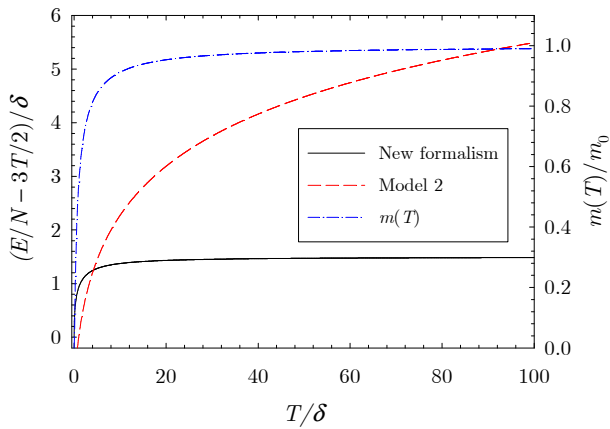}
\caption{(Color online) Comparison between the corrections to the energy for the new formalism and Model 2 as a function of $T/\delta$. We used $T^*/\delta=1$. The evolution of $m(T)/m_0$ is also presented.}
\label{fig11}
\end{figure}

We can now compare this last energy formula with the energy obtained within Models 1 and 2. For the Model 1, where the expression for the energy is preserved, we simply have the standard expression $E/N=3T/2$ while for the Model 2, where the expression for the entropy is preserved, the energy takes the form (remember that in this formalism $\beta=1/T$)
\begin{eqnarray}
	\frac{E}{N}&=&\frac{3T}{2}-\int_{\beta^*}^{\beta}\overline{\partial_{\beta} H}|_{\beta=\nu} d\nu, \nonumber \\
	&=& \frac{3T}{2}+\frac{3}{2}\int_{\beta^*}^{\beta} \frac{\partial_{\nu} m(\nu)}{m(\nu)} \frac{1}{\nu} d\nu, \\
	&=& \frac{3T}{2}+\frac{3\delta}{2}\left[\ln\left(\frac{1+\sqrt{1+4\delta \nu}}{-1+\sqrt{1+4\delta \nu}} \right) \right]_{\nu=1/T^*}^{\nu=1/T}.\nonumber 
\end{eqnarray}
The correction to the energy ($E/N-3T/2$) can be compared for each formalism. Of course, for Model 1, this correction is vanishing; in this case corrections would be associated to the entropy. Consequently, in Fig.~\ref{fig11}, we compare only corrections to the energy obtained with Model 2 and with the new formalism proposed in this paper together with the evolution of the mass as functions of the temperature $T/\delta$.

We notice that the corrections to the energy from Model 2 (and corrections to the entropy from Model 1) are non-local since they involve integrals over some range in temperature, see Eqs.~(\ref{new-entropy}) and ~(\ref{new-energy}). This means that those corrections are still significative in regions where the mass is essentially constant (in this example the corrections are logarithmic in $T$) while the corrections to the energy from the new formalism are essentially localized around the region where the mass depends significantly on the temperature. This is indeed what we expect: If the Hamiltonian does not essentially depend on $T$ over some large interval of temperature, the statistical mechanics in this interval of $T$ should be essentially the same than the standard statistical mechanics.
\end{appendix}

\end{document}